\documentclass[a4paper]{jpconf}

\usepackage{graphicx}
\usepackage{caption}
\usepackage{subfig}
\usepackage{color}
\usepackage{float}
\usepackage{amsmath}
\graphicspath{ {./figures/} }

\newcommand{\leaveout}[1]{}

\begin{document}
\title{Particle-in-Cell Simulations of Plasma Dynamics in Cometary Environment}

\author{Chaitanya Prasad Sishtla, Vyacheslav Olshevsky, Steven W. D. Chien, Stefano Markidis, Erwin Laure}

\address{KTH Royal Institute of Technology, Sweden}

\ead{\{sishtla, slavik, wdchien, markidis, erwinl\}@kth.se}

\begin{abstract}
We perform and analyze global Particle-in-Cell (PIC) simulations of the interaction between solar wind and an outgassing comet with the goal of studying the plasma kinetic dynamics of a cometary environment. To achieve this, we design and implement a new numerical method in the iPIC3D code to model outgassing from the comet: new plasma particles are ejected from the comet ``surface'' at each computational cycle. Our simulations show that a bow shock is formed as a result of the interaction between solar wind and outgassed particles. The analysis of distribution functions for the PIC simulations shows that at the bow shock part of the incoming solar wind, ions are reflected while electrons are heated. This work attempts to reveal kinetic effects in the atmosphere of an outgassing comet using a fully kinetic Particle-in-Cell model.
\end{abstract}

\section{Introduction}
\label{Introduction}
The simulation of interaction between solar wind and cometary plasma is a challenging task because of the peculiarities of the cometary plasma environment. In fact, comets consist of a nuclei composed of a mixture of ices, refractive materials and large organic molecules. These materials sublimate in the form of an outgas due to solar radiation. The extent of sublimation is dependent on the heliocentric distance to the comet. Photoionization and charge exchange with solar wind plasma results in the ionization of sublimated particles leading to the formation of a cometary plasma. The cometary ions are then assimilated by the solar wind through the \textbf{E}$\times$\textbf{B} drift~\cite{holzer1971interaction} and electromagnetic instabilities driven by the cometary ions~\cite{winske1985coupling}. 

The atmosphere of a comet is formed by its outgassing and interaction with solar wind. The innermost region of the comet is the comet nucleus which is an ``icy dirt ball" from which sublimation of particles is happening. The outgassed neutral particles have a density variation $\sim 1/r^2$~\cite{haser1957distribution}. Above the nucleus is the cometary atmosphere (coma) which consists of a ``recombination" region where ionization and recombination of the sublimated particles happen~\cite{goldstein1989observations}. This region exists inside the diamagnetic cavity~\cite{neubauer1993first} which is characterized by a decrease in magnetic field intensity to zero, and with plasma having a radial velocity inside the cavity. The coma is bound by the cometary ionopause. In the innermost regions of the coma where the neutral density is the largest, there are collisions between cometary ions and neutral particles which change the velocity of the cometary ion. For the plasma these collisions act as an ion-neutral friction force. The interaction between outgassed plasma and solar wind strongly depends on the outgassing rate and the direction of the interplanetary magnetic field. Mass loading occurs due to the charge exchange between the newly acquired charged particles and the solar wind. If this exchange is substantial, it leads to the formation of a bow shock~\cite{glassmeier2017interaction}. At the bow shock, picked up particles might exhibit unstable phase space distributions~\cite{coates2009plasma}, leading to kinetic interaction of plasma with solar wind fluctuations and self-generated plasma waves. The goal of this work is to study the kinetic interaction between solar wind and a cometary plasma environment. 


Previous modeling of cometary plasma processes have mainly focused on using either magnetohydrodynamic (MHD) simulations or hybrid simulations, which assume electrons behaving as a fluid and ions are treated as particles. Shou~et~al.~\cite{shou2016new} modeled the coma of the comet using a multi-fluid model with each cometary species treated by separate fluid equations. The simulation used the BATSRUS code~\cite{powell1999solution, toth2012adaptive} and its results are verified against data collected on the 67P/Churyumov-Gerasimenko (CG) comet. MHD models have been developed for simulating comet interactions with the solar wind, and were validated against data collected by the Giotto spacecraft for the Halley comet~\cite{rubin2014comet} and the Rosetta mission for CG~\cite{rubin2015modeled}. Koenders et al.\cite{koenders2015dynamical} used a hybrid model based on the AIKEF code~\cite{muller2011aikef} to simulate the cometary atmosphere taking into account collisions, recombinations and ionizations. Huang et al.~\cite{huang2016four} uses a four-fluid  model to simulate the comet solar wind interaction. However, a complete description of kinetic effects on electrons and ions is only possible via kinetic model. For example, the physical and production mechanisms of non-thermal electron distributions forund in the coma by the Ion and Electron Sensor (IES) onboard Rosetta are still unclear~\cite{madanian2016suprathermal, broiles2016characterizing}. The goal of our work is to use a Particle-in-Cell (PIC) model for studying the kinetic behavior of electrons and ions during the interaction of solar wind with cometary environment.

The first fully kinetic model of a cometary induced magnetosphere was presented by Deca~et~al.~\cite{deca2017electron} using the iPIC3D code\cite{markidis2010multi, peng2015formation}, the authors have shown that the dynamics of ions and electrons in the atmosphere of a weakly outgassing comet could be represented, to first order, in terms of the four-fluid system with solar wind ions, solar wind electrons, cometary ions and cometary electrons each treated by separate fluid equations. Differently from this previous work, we develop an alternative approach for modeling cometary interaction with solar wind using iPIC3D. The key difference of our method is its outgassing mechanism. Instead of creating new plasma particles in the cometary environment at each computational cycle, we inject them from the ``surface'' of the comet according to the physical outgassing rate, i.e., we start from the first principles.

In this paper, we present the simulation approach and the results of our fully kinetic two-dimensional modeling of plasma dynamics in the outgassing comet plasma environment. The paper is organized as follows. Section \ref{Implicit Particle-in-Cell Method} describes the governing equations of implicit  method in the iPIC3D code. Then, Section~\ref{Cometary Plasma Environment} presents the simulation set-up and spatial and temporal scales that are involved in modeling the solar wind interaction with a cometary plasma environment. Section~\ref{Results} presents the simulation results, such as evolution of electron and ions densities, macroscopic flows and velocity distribution functions. Finally, Section~\ref{Discussion and Conclusion} summarizes the results, discusses limitations of this work and outlines future research.
 
\section{Implicit Particle-in-Cell Method}
\label{Implicit Particle-in-Cell Method}
We use the iPIC3D code \cite{markidis2010multi} to model cometary plasma environment from first principles. The PIC method is based on the self-consistent solution of Maxwell's equations and the equations of motion of computational particles (superparticles). Maxwell's equations in iPIC3D are expressed in the second-order (Gaussian or curl-curl) formulation~\cite{jiang1996origin}, with electric field evolution governed by  
\begin{equation}
\nabla^2\mathbf{E}-\frac{1}{c^2}\frac{\partial^2 \mathbf{E}}{\partial t^2}=\frac{4\pi}{c^2}\frac{\partial \mathbf{J}}{\partial t}+4\pi\nabla\rho,
\end{equation}
where $\mathbf{E}$ is the electric field, $c$ is the speed of light, $\mathbf{J}$ is the current density, $\rho$ is the charge density and CGS units are used. 
Once $\mathbf{E}$ is calculated, the magnetic field $\mathbf{B}$ is simply computed from the induction equation:
\begin{equation}
\frac{\partial\mathbf{B}}{\partial t}=-c\nabla\times\mathbf{E}.
\end{equation}

The first two moments of the particle phase space distribution function provide the source terms for the above Maxwell's equations
\begin{equation}
\rho=\sum_s q_s \int f_s d\mathbf{v} \qquad \mathbf{J}=\sum_s q_s \int \mathbf{v} f_s d\mathbf{v}
\label{eq:moments}
\end{equation}
where the index $s$ denotes the plasma species (ions or electrons), $q_s$ is the charge of a particle of the corresponding species, $\mathbf{v}$ is particle velocity, and $f_s(\mathbf{x}, \mathbf{v})$ is the phase space distribution function of species s. The distribution function of species $s$ is then described as a collection of $N_s$ computational particles labeled with index $p$.

The computational particles in the PIC method are used to sample phase space, i.e., each computational particle represents a cloud of real particles in phase space. The functional form of this cloud in iPIC3D is the Dirac's $\delta$ function in the velocity dimensions, and the zeroth order $b$-spline in the spatial dimensions
\begin{equation}
f_s\left( \mathbf{x}, \mathbf{v}, t \right)=b_0\left( \frac{x-x_p}{\Delta x} \right)b_0\left( \frac{y-y_p}{\Delta y} \right)b_0\left( \frac{z-z_p}{\Delta z} \right) \delta(\mathbf{v}-\mathbf{v}_p),
\end{equation}
where $\mathbf{x_p}$ and $\mathbf{v_p}$ refer to the superparticle's position and velocity respectively. The $b_0$ spline is the hat-top function which is nonzero only in the interval $-0.5\leq \zeta\leq 0.5$:
\begin{equation}
b_{0}\left(x\right) = \begin{cases}
            0  & \left| \zeta \right| > 0.5 \\
            1  &  \left|\zeta\right| \leq 0.5.
          \end{cases}
\end{equation}
Particles and grid exchange information by interpolating fields $\mathbf{E}, \mathbf{B}$ (Lorentz force) from the grid to the particles, and by interpolating particle moments such as $\rho, \mathbf{J}$ from the particles to the grid. To evaluate the field $\mathbf{B}$ on the given particle $p$, the following sum is computed over the particle's neighbor grid points $g$
\begin{equation}
\mathbf{B}_p=\sum_g \mathbf{B}_g W\left(\mathbf{x}-\mathbf{x}_p\right).
\end{equation}
To evaluate the moment $\rho$ on the given grid point $g$ the following sum is computed
\begin{equation}
\rho_g=\sum_s^{n_s}\sum_p^{N_s}q_s W\left(\mathbf{x}-\mathbf{x}_p\right),
\end{equation}
and analogously for $\mathbf{J}$, following Eq.~\ref{eq:moments}.
In the above two equations, the key role is played by the interpolation function $W$ which, in case of iPIC3D is the $b$-spline of order $1$, $W\left(x-x_p\right)=b_1\left(x-x_p\right)/\Delta x$. The size of the particle in each spatial dimension corresponds to the grid step in this dimension, $\Delta x$, $\Delta y$ or $\Delta z$.
The $b_1$ spline is nonzero only in the interval $[-1;1]$ where it is a triangular function with maximum $b_1(0)=1$: 
\begin{equation}
b_{1}\left(\zeta\right) = \begin{cases}
            0  & \left| \zeta \right| > 1 \\
            1 - \left| \zeta \right| &  \left|\zeta\right| \leq 1.
          \end{cases}
\end{equation}

Particles move under the effect of an electromagnetic field according to Newton's equation
\begin{equation}
\frac{d\mathbf{x}_p}{dt}=\mathbf{v_p}  \qquad  \frac{d\mathbf{v}_p}{dt}=\frac{q_s}{m_s}\left( \mathbf{E}_p + \frac{\mathbf{v}_p\times \mathbf{B}_p}{c}\right),
\label{eq:newton}
\end{equation}
where $p$ is the particle's index, implying that $\mathbf{E}_p$ and $\mathbf{B}_p$ are interpolated from the grid to the particle's location.

The computational cycle of iPIC3D consists of the following four steps: (1) solve Maxwell's equations discretized on the grid; (2) interpolate new fields and compute the forces acting on each particle; (3) integrate the equations of particle motion \ref{eq:newton}; (4) interpolate new charge and current densities to the grid by integrating particle moments (Eq.~\ref{eq:moments}) in each grid point's volume of control.

The distinctive feature of iPIC3D is the semi-implicit (implicit moment) scheme used to solve Maxwell's equations coupled to the equations of particle motion. In the implicit moment method, an approximate response of the particles to the fields is evaluated at the intermediate time moment at each computational cycle. Approximate values of $\hat{\rho}$ and $\hat{\mathbf{J}}$ are estimated from the Taylor expansion of the interpolation function $W$. This coupling of particle and field equations makes multiple interpolations between the grid and the particles at each computational cycle unnecessary, and dramatically reduces the compute cost compared to the fully implicit method~\cite{markidis2011energy}. This approximate solution, however, has been proven stable and accurate when the stability conditions are satisfied~\cite{brackbill2014multiple}. The stability conditions for implicit PIC methods are much more relaxed compared to the explicit PIC codes which must resolve the smallest physically meaningful scales, such as the Debye length in space and electron plasma frequency in time. Due to relaxed stability restrictions, iPIC3D is capable of modeling fluid-scale processes while retaining the full kinetic description of plasma. 

\section{Cometary Plasma Environment}
\label{Cometary Plasma Environment}

The cometary plasma environment is formed as a result of the sublimation of heavy particles from the nucleus and their interaction with the solar wind. The sublimated particles are ionized by the extreme ultraviolet radiation and undergo inelastic collisions with plasma. The collisions account for the ion-neutral friction and electron cooling. This ionized part of the cometary atmosphere interacts with the incoming solar wind, forming plasma boundaries such as bow shock and ionopause~\cite{nilsson2015birth}. The basic scales and properties of some comets are shown in Table~\ref{comet-basic-scales}. The scales for comets 1P Halley and 26P GS are reported at heliocentric distances of 0.89 AU and 1.01 AU respectively at the time of encounter during the Giotto mission~\cite{krankowsky1986situ, johnstone1993observations}: scales for 67P CG are reported before equinox at approximately 3 AU~\cite{hansen2016evolution}. In the simulation presented outgassing is performed by injecting cometary ions and electrons instead of a neutral gas.

\begin{table}[h!]
\centering
	\begin{tabular}{l|l|l|l}
		\textbf{Comet} & \textbf{Mean Diameter} & \textbf{Production Rate}($\times10^{27}$s$^{-1}$) & \textbf{Ion Gyroradius} \\ \hline
		1P Halley & 11km & 690 & 10000km\\ 
		67P CG & 4km & 0.2 & 1000km\\ 
		26P GS & 2.6km & 7.5 & 4000km\\
	\end{tabular}
	\caption{Mean diameter, production rates and cometary ion gyroradius for comets 1P/Halley, 67P/Churyumov-Gerasimenko and 26P/Grigg-Skjellerup~\cite{coates2009plasma}~\cite{glassmeier2007rosetta}~\cite{hansen2016evolution} are shown. The production rate refers to the gas production rate due to sublimation processes which differ for each comet~\cite{coates2009plasma}.}
	\label{comet-basic-scales}
\end{table}

\subsection{Simulation Setup}
The iPIC3D simulation is carried out in a reduced 2D3V configuration, i.e., two spatial coordinates ($x$, $z$), and three coordinates for velocity, currents, electromagnetic fields, etc.. In iPIC3D, the simulation parameters are normalized with respect to ion scales as shown in Table~\ref{simulation-scales}, where $\omega_{pi}$ is the ion plasma frequency and $d_i = c/~\omega_{pi}$ is the ion skin depth. The comet is placed at the center of the simulation domain measuring 50$d_i$$\times$100$d_i$, with solar wind injection from the left boundary at $x = 0$. The temperatures of solar wind ions and electrons are taken to be 7eV and 10eV respectively. The simulation assumes cometary and solar wind ions to be of equal mass to ensure faster assimilation of cometary ions into the solar wind plasma~\cite{omidi1987kinetic}. A reduced mass ratio of ${m_i}/{m_e} = 64$ is taken to ensure numerical stability, while retaining the scale separation between electron and ion dynamics~\cite{lapenta2010scales}. The interplanetary magnetic field (IMF) is $\textbf{B}_{IMF} = (0, 0, 0.01)~{m_i \omega_{pi} ^{-1}}{e^{-1}}$ where  $e$ is the elementary charge. Together with the IMF, a convective electric field $\textbf{E} = -\mathbf{v_0} \times \textbf{B}_{IMF}$ is initially imposed at $x = 0$, where $\mathbf{v_0}$ is the solar wind bulk velocity. Inflow boundary condition is imposed at $x = 0$ while other boundaries have outflow boundary conditions as discussed in ~\cite{peng2015formation,peng2015kinetic}. We note that the solar wind is super-sonic and super-Alfv\'enic.

\begin{table}[h!]
	\centering
	\begin{tabular}{l|l}
		\textbf{Simulation Parameters} & \textbf{Normalized Values}\\ \hline
		$\rho_{0e}$ , $\rho_{0i}$ & 1, 1\\ 
		$L_x$, $L_z$  & 50~$d_i$, 100~$d_i$\\ 
		$N_x$, $N_z$ & 400, 800 \\ 
		$dx$, $dz$ & 0.125~$d_i$, 0.125~$d_i$ \\
		$(e/m)_i$ & 1\\
		$(e/m)_e$ & -64\\ 
		Radius of comet & 1~$d_i$ \\ 	
		Time step of simulation & 0.15 /~$\omega_{pi}$\\
		Electron thermal velocity & (0.045~$c$, 0.045~$c$, 0.045~$c$)\\
		Ion thermal velocity & (0.0063~$c$, 0.0063~$c$, 0.0063~$c$)\\
		Solar wind bulk velocity & (0.02~$c$, 0, 0)\\
	\end{tabular}
	\caption{Main parameters of the simulation are shown. $n_{0e}$ and $n_{0i}$ are the electron and ion density. $L_x$, $L_z$ are the box lengths in $x$ and $z$ directions respectively. Similarly $N_x$, $N_z$ are the number of cells in $x$ and $z$ directions, and $dx$, $dz$ are the spatial step lengths in $x$ and $z$ directions respectively. All values are normalized with respect to ion scales ($\omega_{pi}$, $d_i$).}. 	\label{simulation-scales}
\end{table}

The main novelty of this paper is the development of a new numerical method to implement particle injection by an emitting boundary, rather than injecting particles throughout the atmosphere. Outgassing is performed by injecting new plasma particles from the ``surface'' of the comet. A spherical shell $5dx$ thick is defined inside the comet and acts as buffer zone similarly to buffer zones created for enabling open boundary conditions~\cite{peng2015formation}. At each time step, new computational particles are created randomly and uniformly distributed inside this spherical shell with particle density $n_{0e} = n_{0i} = 1.0$. In this work, we inject 200 particles per time step. Each new particle receives a thermal speed with a random Gaussian distribution, as specified in Table~\ref{simulation-scales}, plus a radial outgassing velocity of $0.045~c$. The radial outgassing velocity is chosen in a manner to ensure that the ejected cometary ions are fast enough to penetrate farther into the simulation domain within the time scales of the simulation. The strength of outgassing is controlled by the thermal velocity of the outgassed plasma. In the simulation thermal energy of cometary plasma is kept comparable to solar wind plasma to signify strong outgassing~\cite{omidi1987kinetic}. After creation, the particle position is advanced by one time step. If this particle remains inside the comet, it is deleted. Otherwise, it is retained and enters the simulation domain. 

\section{Results}
\label{Results}
In this section, we first show the macroscopic interaction of the solar wind with the cometary plasma environment, showing the electron and ion densities and flows. We then present the distribution functions revealing the microscopic dynamics of electrons and ions.

\subsection{The Formation of the Cometary Plasma Environment}

\begin{figure}[h]
	\centering
	\includegraphics[width=0.8\textwidth]{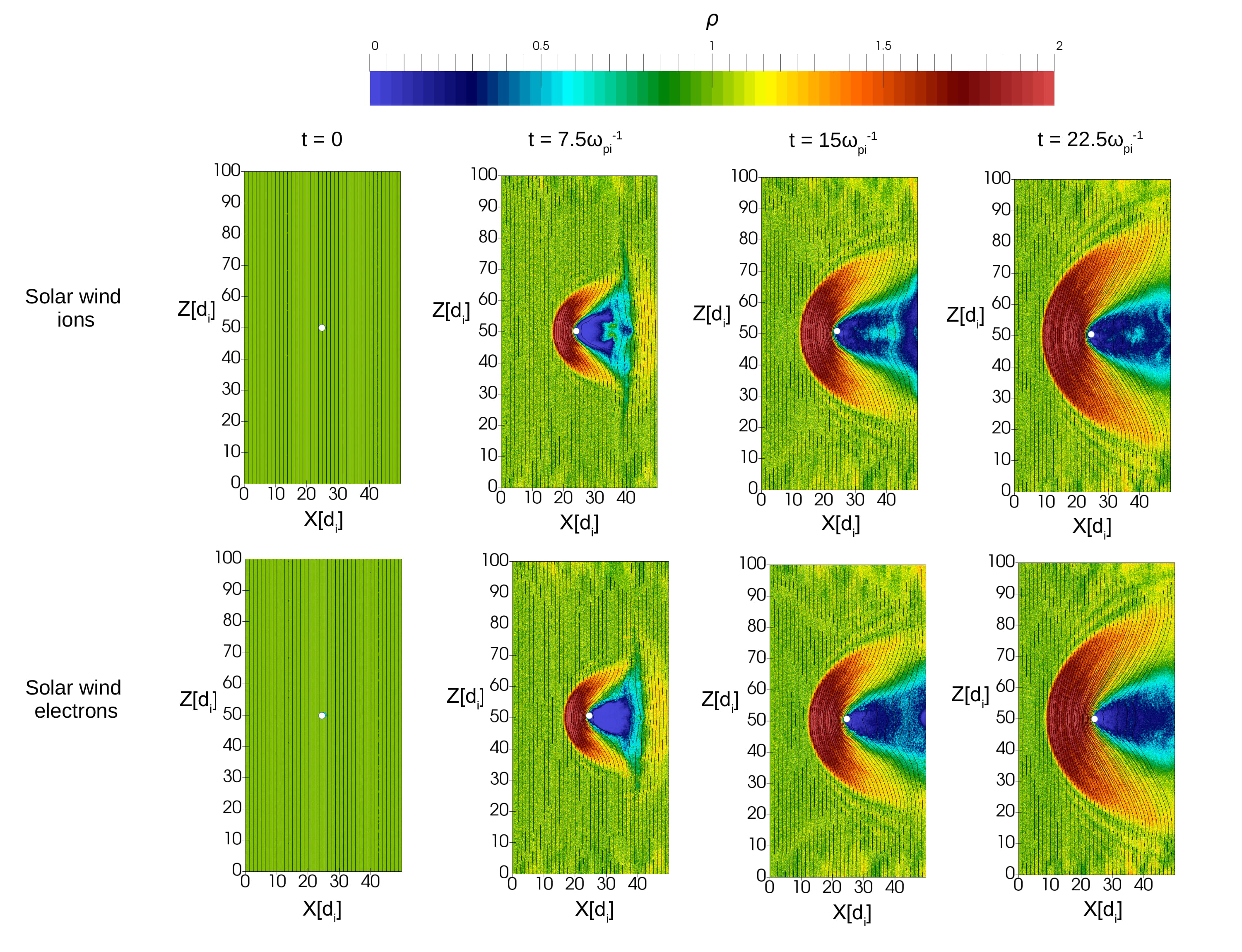}
	\caption{The evolution of solar wind ion and electron species as the simulation progresses is shown in different panels. The simulation starts at $t=0$ and ends at $t = 22.5 \omega_{pi} ^{-1}$.} 
	\label{densityEvol}
\end{figure}

\begin{figure}[h]
	\centering
	\includegraphics[width=0.8 \textwidth]{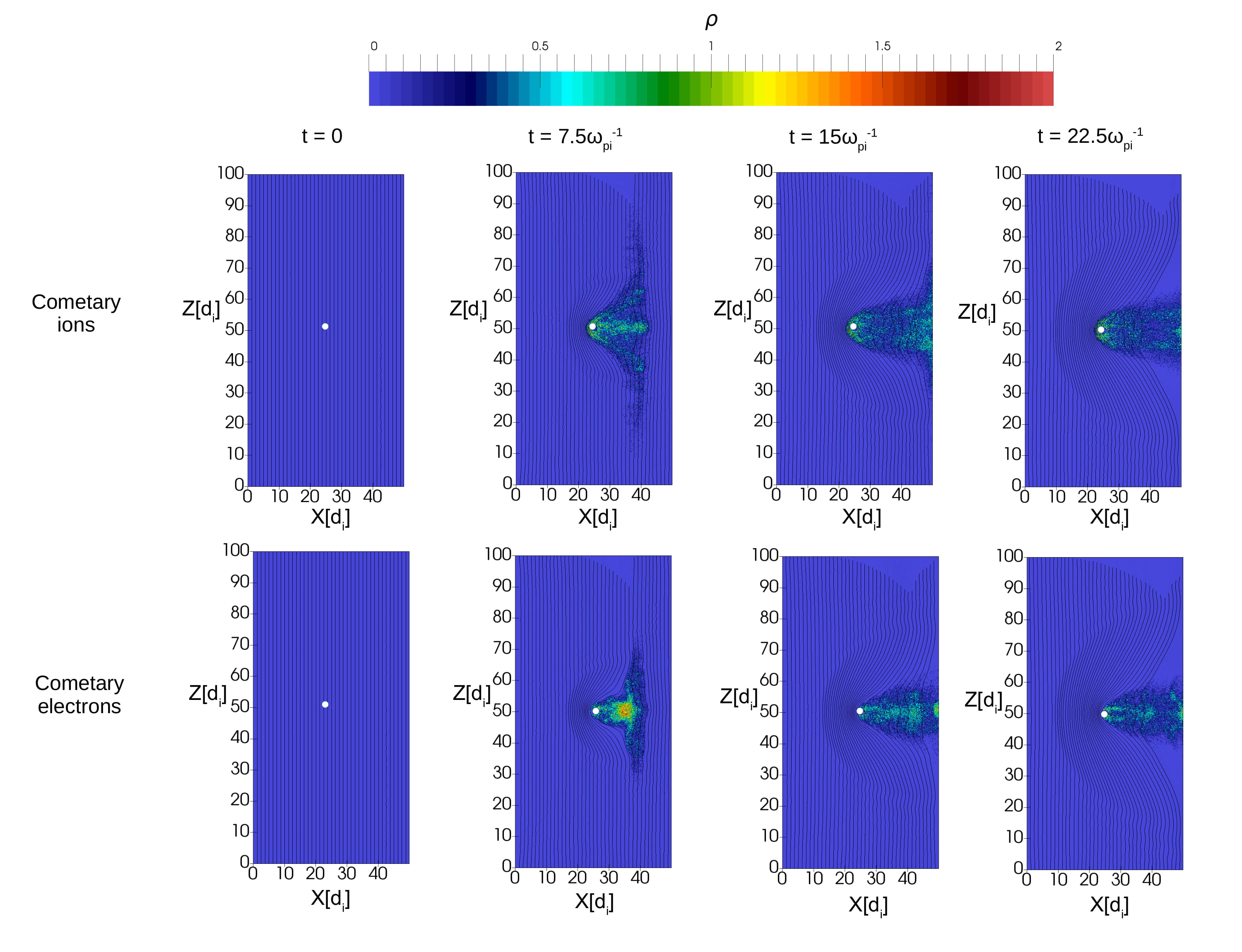}
	\caption{The evolution of cometary ion and electron species as the simulation progresses is shown in different panels. The simulation starts at $t=0$ and ends at $t = 22.5 \omega_{pi} ^{-1}$.} 
	\label{cometEvol}
\end{figure}

In order to create the cometary plasma environment, we initialize the simulation box with a uniform plasma, inject plasma at $x=0$ and at the comet surface. Figures~\ref{densityEvol} and ~\ref{cometEvol} show the contour plots of the densities of solar wind and cometary species superimposed with magnetic field lines at different times during the simulation. The empty circle in the middle of the domain shows the cometary body. When the solar wind flow reaches the comet ($t\omega_{pi} = 7.5$) the formation of a bow shock begins, seen in the density enhancement (roughly twice the initial density) in front of the comet in Figure~\ref{densityEvol}. In Figure \ref{cometEvol} the cometary ions and electrons are seen to be quickly picked up by the incoming solar wind, and are accelerated tailward via the \textbf{J}$\times$\textbf{B} force~\cite{whipple2012physics}. The solar wind electron and ion density in the wake of the comet along the $z = 50 d_i$ line starts decreasing while the cometary electron and ion densities starts increasing. As the simulation progresses and solar wind flow propagates further, the bow shock obtains its classical bell shape, and the low-density solar wind region and high-density cometary region in the wake extends towards the outflow boundary of the simulation domain. The simulated global structure of the solar wind-comet interaction is in accordance with results previously reported by hybrid and PIC simulations~\cite{koenders2015dynamical}. 

\subsection{Cometary Bow Shock}
\begin{figure}[h]
	\centering
	\includegraphics[width=0.8\textwidth]{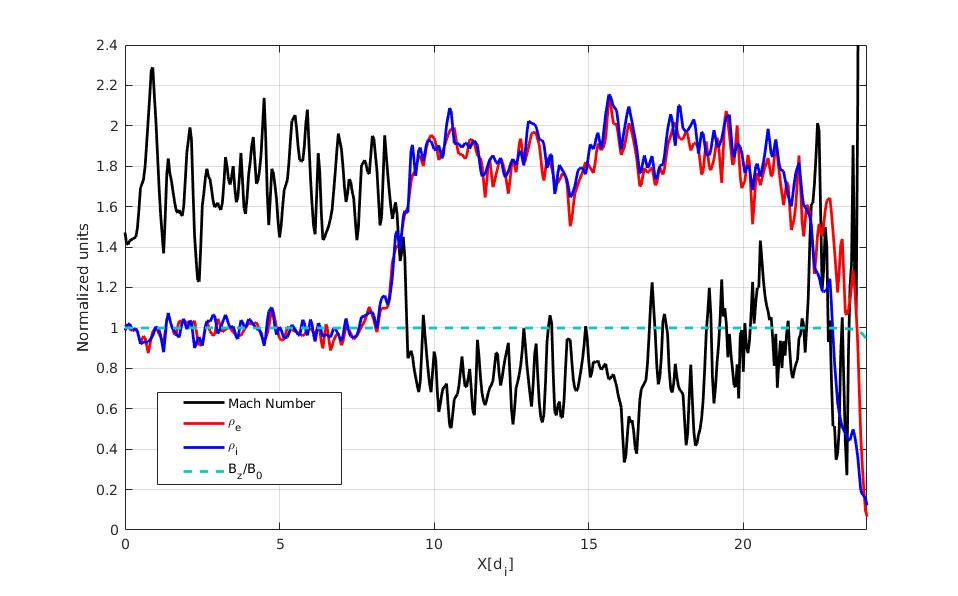}
	\caption{The fast magnetosonic Mach number, $z$ component of the magnetic field normalized to the IMF intensity, and electron and ion densities at time $t = 22.5~\omega_{pi} ^{-1}$.}
	\label{machNum}
\end{figure}

Our simulations show that the interaction of the solar wind and cometary plasma results in the formation of a bow shock. In order to investigate the location and thickness of the bow shock a line plot profiling the magnetosonic Mach number $M=u/(v_A^2+c_{s}^2)$ can be used to observe the transition from super-magnetosonic to sub-magnetosonic flow. Here $u$ is the total electron bulk speed, $v_A$ is the Alfv\'en speed and $c_s = \sqrt{\gamma(T_e/m_i)}$ is the ion sound speed. Figure~\ref{machNum} shows a line plot at $z=L_z/2$ from $x = 0$ to $x = 24 d_i$ of the $z$ component of the magnetic field normalized to the IMF intensity, electron and ion densities, and the fast magnetosonic Mach number calculated using the total electron bulk speed at time $t = 22.5 \omega_{pi} ^{-1}$. The solar wind electron and ion densities overlap indicating very low extent of charge separation, but they both increase by a factor of $2$ after the shock. The bow shock is located at $x \approx 8d_i$, and has a thickness of $\sim 1d_i$.

\begin{figure}[h]
	\centering
	\includegraphics[width=0.8 \textwidth]{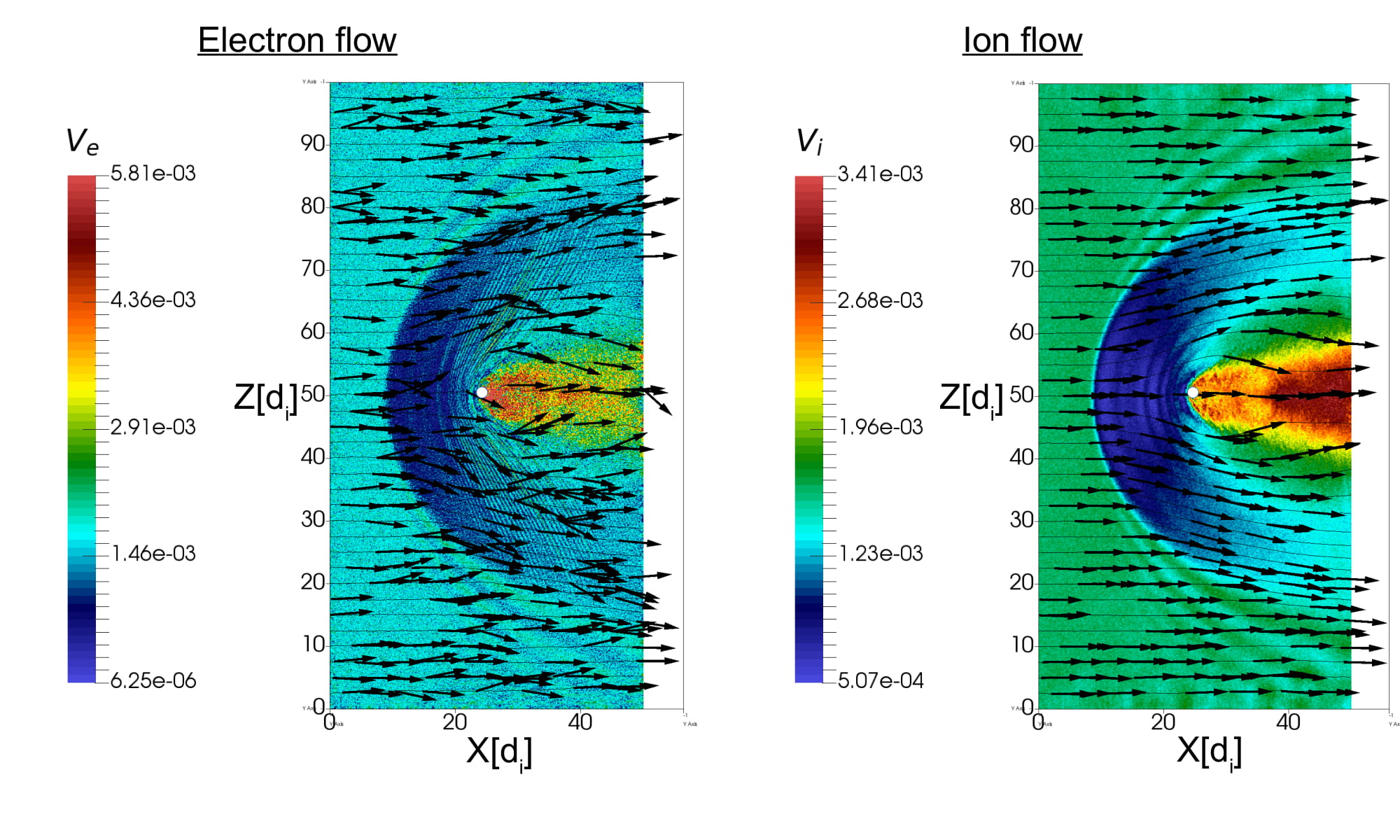}
	\caption{Solar wind electron and ion flows are shown. The figure presents the contour maps of the flow intensities, over which a quiver plot of the flow directions are shown.}
	\label{velocityFlow}
\end{figure}

Figure \ref{velocityFlow} displays the contour plot of solar wind electron and ion flows in normalized units superimposed with a quiver plot showing the flow vectors at random points. Both the solar wind electrons and ions which enter the cometary environment are deflected in the $z$ direction as they move around the comet.


\begin{figure}[h]
	\centering
	\includegraphics[width=0.8 \textwidth]{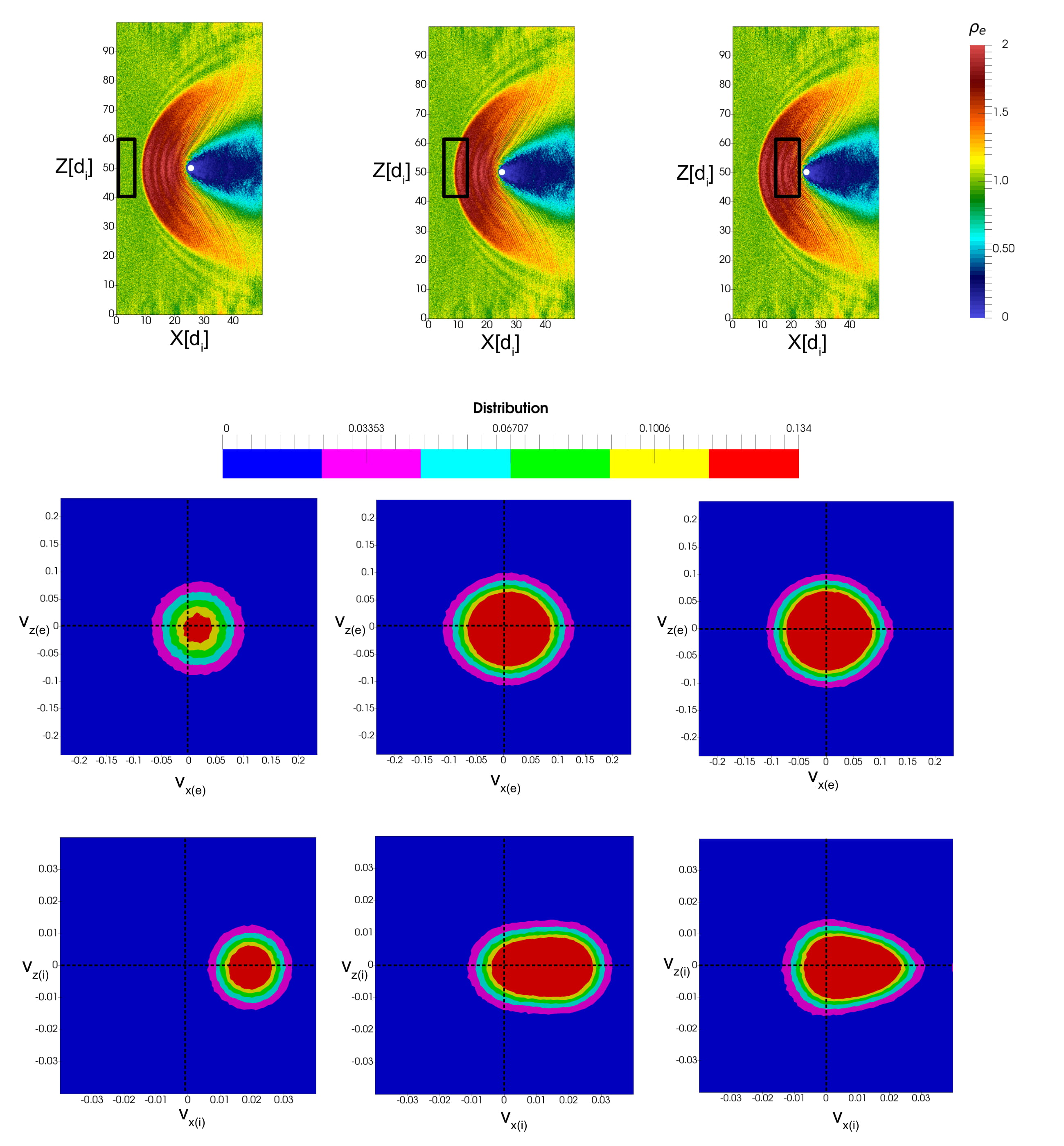}
	\caption{The solar wind electron and ion velocity distribution function in three distinct regions are shown. The first region is present completely in the solar wind, the second region is partly the solar wind and partly the coma while the third region is present completely inside the coma.}
	\label{vdf}
\end{figure}

An analysis of the distribution functions show the details of the microscopic dynamics of solar wind electrons and ions in proximity of the bow shock. We study particle distribution functions in the regions: upstream, bow shock front and downstream. Figure~\ref{vdf} shows the contour plots for solar wind electron and ion velocity distributions in three regions that are highlighted in the upper panels. The first region represents the pristine solar wind; the second region represents an interface between the solar wind and the coma; the third region is immersed into the coma. The solar wind flows in from the left boundary with a bulk speed of $0.02c$. 

The top panel of Figure~\ref{vdf} shows the electron distribution function. Upstream the bow shock in the pristine solar wind, the velocity distribution is purely Maxwellian and centered around $\mathbf{v}=\left(0.02c, 0, 0\right)$. As we move into the transition region between the solar wind and the coma, we see that the incoming electrons are heated especially in the $z$ direction. 

The bottom panel of Figure~\ref{vdf} shows the ion distribution function in the three regions. The distribution function in proximity of the bow shock (bottom center panel in Figure~\ref{vdf}) is particularly interesting: crossing the bow shock, we can observe that the distribution function elongates towards the $-v_x$ axis reaching negative $v_x$ velocity. The ion distribution function shows reflected ions at the bow shock. This is possibly due to the enhanced magnetic field at the bow shock~\cite{koenders2015dynamical}.

\section{Discussion and Conclusion}
\label{Discussion and Conclusion}
We have designed and developed a new fully-kinetic electromagnetic model of an outgassing comet interacting with the solar wind. We have implemented an empirical method for injection of cometary particles in the implicit PIC code iPIC3D: outgassing particles enter the simulation domain from the ``surface'' of the comet with the predefined radial velocity and a random thermal speed. These particles interact with the impinging solar wind and form a well-defined coma, tail, and a bow shock where the interplanetary magnetic field piles up, while the incoming solar wind flow is deflected and reflected. The overall structure of the simulated outgassing comet presented in this paper corresponds to observations and previous hybrid and PIC simulations~\cite{koenders2015dynamical}. The time needed to establish a stationary flow of the solar wind through the simulation domain is comparable to the solar wind transition time.

The incoming flow of the solar wind plasma is deflected in the direction parallel to the magnetic field (perpendicular to the solar wind speed). Closer to the comet, injected electrons mix with the solar wind and heat up, forming a broader distribution function. Solar wind ions mix with the outgassing particles and are reflected at the bow shock due to the enhanced magnetic field, forming an elongated velocity distribution in the $v_x$ direction. This result is consistent with observations and previous simulations. Overall, our new model is promising for more advanced kinetic studies of the cometary atmospheres. 

There are two main limitations to this work. First, we do not model collisions in iPIC3D. A collision model in iPIC3D would allow us to describe the cometary environment with high outgassing rates at higher accuracy. Second, our simulations are in a two dimensional space. This has an effect on the overall exchange of momentum and energy between solar wind and cometary plasma. Our future work will address these two limitations by including a collision model for plasma in iPIC3D and extending the simulations to three dimensions, potentially using iPIC3D and MHD-EPIC~\cite{toth2016extended, daldorff2014two} model. In addition, we plan to complete a parametric study for various solar wind velocities, IMF directions, and the comet outgassing rates.

\section*{Acknowledgments}
This work was funded by Swedish Research (VR) Council project 2017-04508. Experiments were performed on resources provided by the Swedish National
Infrastructure for Computing (SNIC) at PDC Center for High Performance Computing.

\section*{References}
\bibliographystyle{iopart-num}
\bibliography{CometSim}


\end{document}